\documentclass[sn-apa]{sn-jnl}

\usepackage{amsmath} 
\usepackage{amssymb}
\usepackage{amsthm}
\usepackage{tikz}
\usepackage{algorithm}
\usepackage{arevmath}     
\usepackage[noend]{algpseudocode}
\usepackage{program}
\usepackage{graphicx}
\usepackage{booktabs}

\jyear{2022}%

\theoremstyle{thmstyleone}%
\newtheorem{theorem}{Theorem}
\theoremstyle{thmstyletwo}%
\newtheorem{variant}{Variant}%
\newtheorem{application}{Application}%

\theoremstyle{thmstylethree}%
\newtheorem{definition}{Definition}%

\begin{document}

\title[Article Title]{Article Title}

\title{Quantum Computing for a Profusion of Postman Problem Variants
}

\author*[1,2]{\fnm{Joel E.} \sur{Pion}}\email{joelepion@ucsb.edu}

\author[3]{\fnm{Christian F. A.} \sur{Negre}}\email{cnegre@lanl.gov}

\author*[1]{\fnm{Susan M.} \sur{Mniszewski}}\email{smm@lanl.gov}

\affil[1]{\orgdiv{Computer, Computational, and Statistical Sciences Division}, \orgname{Los Alamos National Laboratory}, \state{NM}, \country{USA}}

\affil[2]{\orgdiv{Mathematics Department}, \orgname{University of California, Santa Barbara}, \state{CA}, \country{USA}}

\affil[3]{\orgdiv{Theoretical Division}, \orgname{Los Alamos National Laboratory}, \state{NM}, \country{USA}}


\abstract{In this paper we study the viability of solving the Chinese Postman Problem, a graph routing optimization problem, and many of its variants on a quantum annealing device. Routing problem variants considered include graph type, directionally varying weights, number of parties involved in routing, among others. We put emphasis on the explanation of how to convert such problems into quadratic unconstrained binary optimization (QUBO) problems, one of two equivalent natural paradigms for quantum annealing devices. We also expand upon a previously discovered algorithm for solving the Chinese Postman Problem on a closed undirected graph to decrease the number of constraints and variables used in the problem. Optimal annealing parameter settings and constraint weight values are discussed based on results from implementation on the D-Wave 2000Q and Advantage. Results from classical, purely quantum, and hybrid algorithms are compared.}




\date{Received: date / Accepted: date}

\maketitle

\section{Introduction}
\label{intro}
Quantum annealing exploits the quantum-mechanical effects of superposition, entanglement, and tunneling to explore the energy landscape in an efficient manner \cite{Lanting2014,annealingbasics} when 
sampling from energy-based models. 
NP-hard combinatorial optimization problems are formulated as either an Ising model or quadratic unconstrained binary optimization (QUBO) problem that can be run on a D-Wave quantum annealer (QA). The Ising model objective function is 

\begin{equation}
    O({h,J,s}) = \sum\limits_{i}h_{i}s_i + \sum\limits_{i<j}J_{ij}s_is_j,
    \label{eq:ising}    
\end{equation}
where $s_{i} \in$ $\{-1,+1\}$ are the spin variables, 
while $h_{i}$ and $J_{ij}$ are, respectively, biases on, and strengths between spins. 

Quantum computers use qubits to encode information. Their behavior is governed by the laws of quantum mechanics. This allows a qubit to be in a “superposition” state which means it can be both a “$-1$” and a “$+1$” at the same time. An outside event causes it to collapse into either. The annealing process results in a low-energy ground state, $g$, which consists of an Ising spin for each qubit.

While solving a combinatorial optimization problem, QAs are typically limited by the number of variables which can be represented and embedded in the hardware graph topology. During the embedding process each logical variable maps to a chain of qubits.
The D-Wave 2000Q QA uses a Chimera topology with more than $2000$ qubits and more than $6000$ couplers. Each qubit is connected to $6$ others. This allows a fully connected graph (or clique) of $64$ nodes 
(or variables) to be embedded in the sparse Chimera graph. The newer D-Wave Advantage uses a Pegasus topology with over $5000$ qubits and more than $35,000$ couplers. Each qubit connects to $15$ other qubits, and the largest embeddable clique size is $177$ \cite{McGeoch}. 

QAs proved to be useful for solving NP-hard optimization problems such as those involved in graph theory \cite{GP2017,CD2020,Mniszewski2021} and machine learning \cite{OMalley2018,Dixit2021} among others. 
The QUBO formulation is most commonly used for optimization problems. The objective function is

\begin{equation}
    O({Q,x}) = \sum\limits_{i}Q_{ii}x_i + \sum\limits_{i<j}Q_{ij}x_ix_j,
    \label{eq:qubo}    
\end{equation}
where $x_{i} \in$ $\{0,1\}$ encodes the inputs and results.
The symmetric matrix, $Q$, is formulated such that the weights on the diagonal correspond to the linear terms, while the off-diagonal weights are the quadratic terms.
Ising and QUBO models are related through the
transformation $s = 2x - 1$. 

Constraints on current D-Wave architectures include limited precision and range on weights and strengths, sparse connectivity, and number of available qubits. These constraints impact both the size of the problems that can be run and the solver performance. A QUBO matrix is mapped onto the hardware using an embedding algorithm such as \emph{minorminer}~\cite{embedding}. 
A hybrid quantum-classical approach is required when the number of problem variables is too large to run directly on the D-Wave hardware. In that case, the quantum-classical \emph{qbsolv} sampler is used \cite{Booth}.


This paper presents a routing problem known as the Chinese Postman Problem (CPP) as well as many of its variants in the context of finding solutions with a QA. In Section \ref{Preliminaries}, we briefly give an overview of
the history of the CPP as well as some definitions necessary to define the problem.
In Section \ref{variants}, we define
many variants of the Postman Problem
along with potential applications.
Next, in section \ref{Undirected Explanation},
a CPP QA algorithm first put forth by \cite{Siloi}
for solving 
the Closed Undirected CPP is discussed with potential modifications.
Then, we introduce a novel algorithm for using a QA to solve a large class of CPP variants in sections \ref{BGC} and \ref{even more general}.
Finally, we show results from our implementations and discuss 
observations from our experiments in sections \ref{results} and \ref{discussion} respectively.

\section{Preliminaries}\label{Preliminaries}

\subsection{History}
\label{History}
The CPP was first posed as a combinatorial optimization problem by the then lecturer at Shandong Normal University, Mei-Gu Guan, in 1960. 
At that time China was trying to modernize itself as a country and
mathematicians were encouraged to work on real-world applications. The original phrasing of the CPP is as follows: 

``\textit{A postman has to deliver letters to a given neighborhood. He needs to walk through all the streets in the neighborhood and back to the post-office. How can he design his route so that he walks the shortest distance? \cite{Grotschel}}"

The modern phrasings of the question are varied as different factors and applications are taken into consideration. The unifying factor across these variants are that they are routing problems framed as combinatorial optimization problems over a graph structure.

\subsection{Graph Algorithm Terminology}
\label{Graph Definitions}

The following are definitions about graphs and objects used in graph algorithms discussed in this paper. The following definitions will be used throughout the paper \cite{graphtextbook}.

\begin{definition}
\label{definition: graph}
(Graph) A graph, $G$, is a triple ($V$, $U$, $D$) where 
$V\subset\mathbb{N}$ is a non-empty finite subset,  
$U\subset\{([a,b],c) \vert a,b\in V, c\in(\mathbb{R}^+)^2\}$
with $c = [W_{a,b}, W_{b,a}]$, and $D\subset\{((a,b),W_{a,b}) \vert a,b\in V, W_{a,b}\in\mathbb{R}^+\}.$ We shall refer to $V$ 
as vertices, $U$ as undirected edges, and $D$ as directed edges. Undirected edges and directed 
edges are labeled as $[a,b], (a,b)$, respectively, while $W_{a,b}$ is the weight of the edge from vertex $a$ to vertex $b$. We refer to $E = U\bigcup D$ as the edges.

Note $(*,...,*)$ is used to denote an ordered tuple and $[*,...,*]$ is used to denoted an unordered tuple.
\end{definition}

\begin{definition}
(Vertex Adjacency) In a graph, $G$, vertex $a\in V$ is said to be adjacent to vertex $b\in V$ if there exists some $W_{a,b}, W_{b,a}\in\mathbb{R}^+$ so that $([a,b],[W_{a,b},W_{b,a}])\in U$ or there exists some $W_{a,b}\in\mathbb{R}^+$ so that ${((a,b),W_{a,b})\in D}$.
\end{definition}

\begin{definition}
(Edge Adjacency) In a graph, $G$, the edge labeled $[a,b]$ or $(a,b)$ is adjacent to the edge labeled $[c,d]$ or $(c,d)$ if the edges may be written, up to reordering of unordered tuples, so that $b = c$.
\end{definition}

\begin{definition}
(Walk) A walk of length $n$ in a graph, $G$, is a tuple of length $n+1$, $(v_0,...,v_n)$, where $v_i$ is a vertex adjacent to $v_{i+1}$ for all $i$. 
\end{definition}

\begin{definition}
(Open/Closed Walk) A walk in a graph is a closed walk if the first and last vertex in the walk are the same. Otherwise the walk is called an open walk.
\end{definition}

\begin{definition}
(Walk Weight) The walk weight of a walk is the sum of all the weights of the edges traversed in the walk. Given a walk $(v_0,...,v_n)$, the walk weight is $\sum\limits_{i=0}^{n-1}W_{v_i,v_{i+1}}$.
\end{definition}

\begin{definition}
(Trail) A trail is a walk for which no edge is repeated within the walk.
\end{definition}

\begin{definition}
(Circuit) A circuit is a closed trail. 
\end{definition}

\begin{definition}
(Eulerian Circuit) An Eulerian circuit is a circuit which includes every edge in the graph.
\end{definition}

\begin{definition}
(In/Out-Degree) A vertex, $v$, in a graph, $G$, has in-degree equal to the number of vertices adjacent to $v$, and out-degree equal to the number of vertices $v$ is adjacent to. In other words, the in-degree of the vertex $v$ is the number of edges which end in $v$, up to reordering of unordered tuples. The out-degree is similar in reverse.   
\end{definition}

\begin{definition}
(Degree) A vertex, $v$, in an undirected graph, $G$, has degree equal to its in-degree and out-degree.  
\end{definition}

\begin{definition}
(Strongly Connected) A graph $G$ is said to be strongly connected if for every pair of vertices, $a,b\in V$, there exists a walk in $G$ from vertex $a$ to vertex $b$.
\end{definition}

\begin{definition}
(Partially Ordered Set) A pair ($X$,$\leq$) such that $X$ is a set, $S\subset X\times X$ (Cartesian product), and $x\leq y$ for $x,y\in X$ if and only if $(x,y)\in S$, is called a partially ordered set if the following hold: 
\begin{enumerate}
    \item $x\leq x$ for all $x\in X$\\
    \item $x\leq y$ and $y\leq x \implies x = y$\\
    \item $x\leq y$ and $y\leq z \implies x\leq z$
\end{enumerate}
\end{definition}

\begin{definition}
(Perfect Pairing) Let $S$ be a finite set with an even number of elements. Then a perfect pairing of $S$ is a collection of subsets of $S$, $A_i$, such that: 
\begin{enumerate}
    \item $\vert A_i\vert$ = 2 for all i\\
    \item $A_i\cap A_j = \emptyset$ for all $i\neq j$\\
    \item $\bigcup A_i = S$
\end{enumerate}
\end{definition}

We shall assume henceforth that all our graphs are strongly connected. 

\section{Methods}
\label{methods}

\subsection{Variants and Applications}
\label{variants}

The CPP is a general term for a wide variety of routing problems. Each variant of the CPP is often created to optimize a specific application
~\cite{Thimbleby,Comaklisokmen}.

\begin{variant}
\label{undirected variant}
(Undirected CPP) Given an undirected graph, $G$, 
find a walk in $G$ which traverses every edge in $G$ with the minimal walk weight.
\end{variant}

\begin{application}
(Neighborhood Pothole Inspection) Imagine one wished to survey the road conditions in a large neighborhood with bidirectional roads. One could represent the neighborhood as an undirected graph with intersections as vertices and the roads as edges and solve the Undirected CPP.
\end{application}

\begin{variant}
\label{directed variant}
(Directed CPP) Given a directed graph, $G$,
find a walk in $G$ which traverses every edge in $G$ with the minimal walk weight.
\end{variant}

\begin{application}
(Downtown Pothole Inspection) Imagine one wished to survey the road conditions of a city's downtown containing only one-way streets. One could represent the downtown area as a directed graph with the intersections as vertices and the roads as edges and solve the Directed CPP.
\end{application}

\begin{variant}
\label{mixed variant}
(Mixed CPP) Given a mixed graph, $G$,
find a walk in $G$ which traverses every edge in $G$ with the minimal walk weight.
\end{variant}

\begin{application}
(Town Pothole Inspection) Imagine one wished to survey the road conditions of an entire town which contained one-way streets, two-way streets (e.g. a highway with lanes), and bidirectional streets (e.g. a residential street with no lanes). One could represent the town as a mixed graph with the intersections as vertices, the one/two-way streets as one/two directed edges, and the bidirectional streets as undirected edges, and solve the Mixed CPP.
\end{application}

The above variants determine what types of graphs need to be considered for the problem, which can drastically effect the computational complexity of the problem. Both the undirected and the directed variants are solvable classically in polynomial time, while the mixed variant is NP-Hard~\cite{Comaklisokmen}.
For any CPP one will need to choose a type of graph to work over as well as 
where the postman will need to start and/or stop their route. When solving the CPP classically, the start/stop choice will change what algorithm is needed~\cite{Thimbleby}.

\begin{variant}
\label{closed variant}
(Closed CPP) Given a graph, $G$, find a walk in $G$ which traverses every edge in $G$ such that the start and stop are on the same vertex, with a minimal walk weight for such a walk.
\end{variant}

\begin{application}
(Tunnel Inspections) Imagine a mine operator wishes to inspect the integrity of the tunnels in her mining operation. The mine has only one entry/exit for its vast and extensive network of tunnels. One could represent the tunnels as edges and the tunnel junctions as vertices and solve the Closed CPP.
\end{application}

\begin{variant}
\label{open variant}
(Open CPP) Given a graph, $G$, find a walk in $G$ which traverses every edge in $G$ with the minimal walk weight.
\end{variant}

\begin{application}
(Museum Cleaning Robot) Imagine a museum wishes to clean their floors using a robot and has two docking stations for the robot start and stop at. The museum wishes to know where to place these docking stations as well as what walk the robot should take so as to clean the museum efficiently. One could represent the rooms as vertices and the hallways between them as edges and solve the Open CPP.
\end{application}

\begin{variant}
\label{endpoint variant}
(Open with Endpoints CPP) Given a graph, $G$, and a starting vertex, $v_1$, and/or a stopping vertex, $v_2$, find a walk in $G$ which traverses every edge in $G$ with the minimal walk weight for such walks. This walk starts at $v_1$ if given and ends at $v_2$ if given.
\end{variant}

\begin{application}
(Botanical Garden Picnic) Imagine you wish to see every part of the Botanical Garden as efficiently as possible. You may wish to just start at the beginning and finish at the exit or you may wish to start at the beginning and finish somewhere in the garden (you don't care where) so as to enjoy a picnic. One could represent the paths as edges and the path junctions as vertices and solve the Open with Endpoints CPP.
\end{application}

Now that all of the required variant choices have been laid out, we will introduce some optional variants to modify the CPP. Inclusion of these variants allow the CPP to be applied in a much wider array of applications. Many of the variants can be applied in conjunction with one another so as to be applicable in an even broader set of use cases.

\begin{variant}
\label{rural variant}
(Rural Postman Problem) Given a graph, $G$, and a subset, $R$, of the edges of $G$, find a walk in $G$ which traverses every edge in $R$ with the minimal walk weight.
\end{variant}

\begin{application}
(Traveling Salesmen) Imagine you were a traveling salesmen who wished to sell your wares in every capital city in every state of the United States of America. One could consider the graph defined with every capital city as a vertex and with an edge (weighted by cost to travel) between every pair of capital cities. Then one could modify this graph by replacing each vertex with two vertices (each with all the original vertices edges) with an edge of weight zero connecting them. Considering only the added edges of weight $0$ as our $R$, one could solve the Rural Postman Problem. 
\end{application}

\begin{variant}
\label{windy variant}
(Windy Postman Problem) Given a graph, $G$, where $W_{a,b}$ may not equal $W_{b,a}$ for undirected edges, as in definition \ref{definition: graph}, find a walk in $G$ which traverses every edge of $G$ with minimal walk weight. 
\end{variant}

\begin{application}
(Injured Hiker) Imagine a rescue team is trying to find an injured hiker on the trails in a mountain range. One could represent the trails as undirected edges and the trail junctions as vertices. One could then account for the differences in difficulties of going uphill versus downhill by assigning different directional weights to the undirected edges and solve the Windy Postman Problem. 
\end{application}

\begin{variant}
\label{k variant}
($k$-Postman Problem With Capacity) Given a graph, $G$, $k\in\mathbb{N}$, and $\{c_1,...,c_k\}\subset(\mathbb{R}\cup\{\infty\})^k$, find $k$ walks in $G$ such that each edge in $G$ is covered by at least one of the $k$ walks, the $i$\textsuperscript{th} walk weight is less than or equal to $c_i$ for $i\in\{1,...,k\}$, and the sum of walk weights are minimized. 
\end{variant}

\begin{application}
(Postal Service) Imagine you were in charge of your local area postal service and had $10$ postal agents/vehicles in your employ. You need to deliver mail to every street in your region in an efficient way, but no postal worker may work more than an 8 hour workday by law. One could represent every street as an edge and every street intersection as a vertex and solve the $k$-Postman Problem With Capacity.
\end{application}

\begin{variant}
\label{service variant}
(Service-Based Traversal Postman Problem) Given a graph, $G$, modify the graph so as to create a duplicate of each edge, without duplicating any vertices. The duplicated edges may have a different weight. Note that each pair of vertices which had only one edge now has two edges between them. All the original edges will be called servicing edges, while all the added edges will be called traversal edges. One should then solve the Rural Postman Problem on all the servicing edges. 
\end{variant}

\begin{application}
(Pipe Repairman) Imagine your were a pipe repairman and you had an extensive network of pipes to repair. It takes you 1 hour to repair $10$ meters of pipe and $10$ minutes to pull your pipe fixing supplies that same distance. One could represent the pipes as edges and the pipe-splitting junctions as vertices and solve the Service-Based Traversal Postman Problem.
\end{application}

\begin{variant}
\label{turning variant}
(Turning Challenge Postman Problem) Given a graph, $G$, and a collection of $3-$tuples in the form (edge-in, edge-out, bonus weight), we shall, with regard to the collection of $3-$tuples, sum the corresponding bonus weights for each instance where edge-in is followed by edge-out in the walk. We will call this sum, extra weight. Find a walk which traverses every edge in $G$ where the sum of the walk weight and extra weight is minimized.
\end{variant}

\begin{application}
(Street Cleaner) Imagine your job was to clean the streets in a North American city where there are lights and stop-signs. In general, it is faster to go right or straight than it is to make a left turn or a u-turn. One could form a collection of $3-$tuples by, for each road, $r$, making $3-$tuples of the form $(r,s,w)$ for each road, $s$, which could follow $r$, with $w$ being the added time it takes to make such a turn. Then one could represent each road as an edge and each intersection as a vertex and solve the Turning Challenge Postman Problem.
\end{application}

\begin{variant}
\label{hierachy variant}
(Hierarchical Postman Problem) Given a graph, $G$, for which the edges, $E$, have a partial ordering, find a Service/Traversal Postman Problem solution constrained by the edges needing to be serviced in an order congruous to the partial order.
\end{variant}

\begin{application}
(Forgotten Packages) Imagine you were a delivery person in a town and yesterday several packages were delivered to the wrong address. Now you must deliver today's packages as well as pick-up and redeliver the misdelivered packages. One could represent roads as edges and the intersections as nodes. Then one could place a partial ordering on the edges so that a street with a package which was misdelivered yesterday must come before the street with the intended destination of the package. One could then solve the Hierachical Postman Problem.
\end{application}

\subsection{Foundations for Solving a Problem on a Quantum Annealing Device}
\label{QUBO foundation}

At its core, QAs are machines which are meant to solve one kind of problem extremely well. Fortunately, that problem is NP-Complete~\cite{Lewis}
and many useful problems may efficiently be converted into an instance of this problem. Ising model and QUBO formulations of the aforementioned problems can be solved on a QA. We will frame the CPP problem as a Polynomial Version QUBO (see definition \ref{Polynomial Version QUBO} below).

\begin{definition}
(Matrix Version: QUBO Problem) Given $Q\in M_n(\mathbb{R})$, find $$min\{\vec{x}^\top Q\vec{x}\}$$
constrained by $$\vec{x} = 
\begin{pmatrix} 
x_1 \\ 
x_2 \\ 
... \\
x_n
\end{pmatrix} \text{ with } x_i\in\{0,1\}\text{ for all } i.$$
\end{definition}

\begin{definition}
\label{Polynomial Version QUBO}
(Polynomial Version: QUBO Problem) Given $q_{ij}\in\mathbb{R}$ for $i,j\in\{1,...,n\}$, find $$min\{\sum\limits_{i = 1}^n\sum\limits_{j = 1}^n q_{ij}x_ix_j\}$$ 
constrained by $x_i\in \{0,1\}\text{ for all }i$.
\end{definition}

Once one realizes that any binary variable $x\in\{0,1\}$ has the property $x^2 = x$~\cite{Glover}, 
the equivalence of the two versions becomes clear upon inspection. Now that we have defined what a QUBO is, we outline the individual steps involved in solving a problem with a QUBO formulation on a QA (see Figure \ref{workflow diagram} ). 

\tikzstyle{decision} = [diamond, draw, fill=blue!20, 
    text width=6em, text badly centered, node distance=3cm, inner sep=0pt]
\tikzstyle{block} = [rectangle, draw, fill=blue!20, 
    text width=5em, text centered, rounded corners, minimum height=4em]
\tikzstyle{line} = [draw, -latex']
\tikzstyle{cloud} = [draw, ellipse,fill=red!20, node distance=3cm,
    minimum height=2em]

\tikzset{every picture/.style={line width=0.75pt}} 
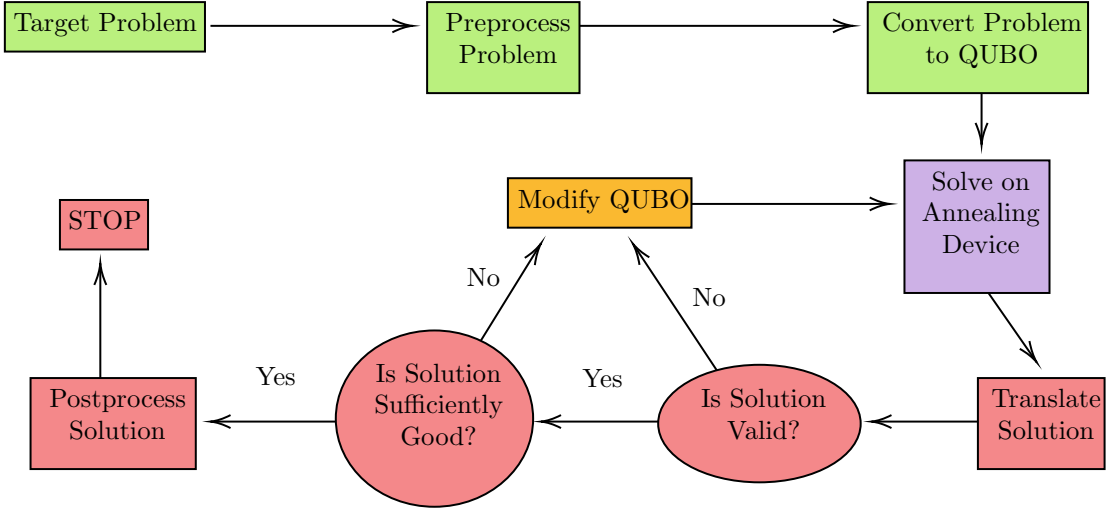
\begin{figure}[h!]
    \centering
    \caption{Quantum Annealing Workflow. The specifics of the green and red portions of the diagram are dependent on the problem one is trying to solve. The green portion is the classical work to prepare the input to the QA, while the red portion is the classical work to prepare the output of the QA. The purple block can be done via quantum annealing, simulated annealing, or a quantum-classical approach like \emph{qbsolv}~\cite{Booth}. The orange block is where one needs to tune the problem's input parameters.
    }
    \label{workflow diagram}
\begin{tikzpicture}[x=0.75pt,y=0.75pt,yscale=-1, xscale=0.918, trim left = 1.5cm]

\draw    (130,30) -- (238,30) ;
\draw [shift={(240,30)}, rotate = 180] [color={rgb, 255:red, 0; green, 0; blue, 0 }  ][line width=0.75]    (10.93,-3.29) .. controls (6.95,-1.4) and (3.31,-0.3) .. (0,0) .. controls (3.31,0.3) and (6.95,1.4) .. (10.93,3.29)   ;
\draw    (330,30) -- (478,30) ;
\draw [shift={(480,30)}, rotate = 180] [color={rgb, 255:red, 0; green, 0; blue, 0 }  ][line width=0.75]    (10.93,-3.29) .. controls (6.95,-1.4) and (3.31,-0.3) .. (0,0) .. controls (3.31,0.3) and (6.95,1.4) .. (10.93,3.29)   ;
\draw    (550,60) -- (550,88) ;
\draw [shift={(550,90)}, rotate = 270] [color={rgb, 255:red, 0; green, 0; blue, 0 }  ][line width=0.75]    (10.93,-3.29) .. controls (6.95,-1.4) and (3.31,-0.3) .. (0,0) .. controls (3.31,0.3) and (6.95,1.4) .. (10.93,3.29)   ;
\draw    (550,160) -- (578.8,198.4) ;
\draw [shift={(580,200)}, rotate = 233.13] [color={rgb, 255:red, 0; green, 0; blue, 0 }  ][line width=0.75]    (10.93,-3.29) .. controls (6.95,-1.4) and (3.31,-0.3) .. (0,0) .. controls (3.31,0.3) and (6.95,1.4) .. (10.93,3.29)   ;
\draw    (550,230) -- (492,230) ;
\draw [shift={(490,230)}, rotate = 360] [color={rgb, 255:red, 0; green, 0; blue, 0 }  ][line width=0.75]    (10.93,-3.29) .. controls (6.95,-1.4) and (3.31,-0.3) .. (0,0) .. controls (3.31,0.3) and (6.95,1.4) .. (10.93,3.29)   ;
\draw    (380,230) -- (312,230) ;
\draw [shift={(310,230)}, rotate = 360] [color={rgb, 255:red, 0; green, 0; blue, 0 }  ][line width=0.75]    (10.93,-3.29) .. controls (6.95,-1.4) and (3.31,-0.3) .. (0,0) .. controls (3.31,0.3) and (6.95,1.4) .. (10.93,3.29)   ;
\draw    (200,230) -- (132,230) ;
\draw [shift={(130,230)}, rotate = 360] [color={rgb, 255:red, 0; green, 0; blue, 0 }  ][line width=0.75]    (10.93,-3.29) .. controls (6.95,-1.4) and (3.31,-0.3) .. (0,0) .. controls (3.31,0.3) and (6.95,1.4) .. (10.93,3.29)   ;
\draw    (70,210) -- (70,152) ;
\draw [shift={(70,150)}, rotate = 90] [color={rgb, 255:red, 0; green, 0; blue, 0 }  ][line width=0.75]    (10.93,-3.29) .. controls (6.95,-1.4) and (3.31,-0.3) .. (0,0) .. controls (3.31,0.3) and (6.95,1.4) .. (10.93,3.29)   ;
\draw    (390,120) -- (498,120) ;
\draw [shift={(500,120)}, rotate = 180] [color={rgb, 255:red, 0; green, 0; blue, 0 }  ][line width=0.75]    (10.93,-3.29) .. controls (6.95,-1.4) and (3.31,-0.3) .. (0,0) .. controls (3.31,0.3) and (6.95,1.4) .. (10.93,3.29)   ;
\draw    (270,200) -- (308.89,141.66) ;
\draw [shift={(310,140)}, rotate = 123.69] [color={rgb, 255:red, 0; green, 0; blue, 0 }  ][line width=0.75]    (10.93,-3.29) .. controls (6.95,-1.4) and (3.31,-0.3) .. (0,0) .. controls (3.31,0.3) and (6.95,1.4) .. (10.93,3.29)   ;
\draw    (410,210) -- (361.16,141.63) ;
\draw [shift={(360,140)}, rotate = 54.46] [color={rgb, 255:red, 0; green, 0; blue, 0 }  ][line width=0.75]    (10.93,-3.29) .. controls (6.95,-1.4) and (3.31,-0.3) .. (0,0) .. controls (3.31,0.3) and (6.95,1.4) .. (10.93,3.29)   ;

\draw  [fill={rgb, 255:red, 186; green, 240; blue, 126 }  ,fill opacity=1 ]  (18,18) -- (127,18) -- (127,43) -- (18,43) -- cycle  ;
\draw (21,22) node [anchor=north west][inner sep=0.75pt]   [align=left] {Target Problem};
\draw  [fill={rgb, 255:red, 186; green, 240; blue, 126 }  ,fill opacity=1 ]  (248,18) -- (331,18) -- (331,64) -- (248,64) -- cycle  ;
\draw (251,22) node [anchor=north west][inner sep=0.75pt]   [align=left] {\begin{minipage}[lt]{54.31pt}\setlength\topsep{0pt}
\begin{center}
Preprocess\\Problem
\end{center}

\end{minipage}};
\draw  [fill={rgb, 255:red, 186; green, 240; blue, 126 }  ,fill opacity=1 ]  (488,18) -- (608,18) -- (608,64) -- (488,64) -- cycle  ;
\draw (491,22) node [anchor=north west][inner sep=0.75pt]   [align=left] {\begin{minipage}[lt]{79.26pt}\setlength\topsep{0pt}
\begin{center}
Convert Problem\\to QUBO
\end{center}

\end{minipage}};
\draw  [fill={rgb, 255:red, 205; green, 177; blue, 230 }  ,fill opacity=1 ]  (508,98) -- (587,98) -- (587,165) -- (508,165) -- cycle  ;
\draw (511,102) node [anchor=north west][inner sep=0.75pt]   [align=left] {\begin{minipage}[lt]{50.93pt}\setlength\topsep{0pt}
\begin{center}
Solve on\\Annealing \\Device
\end{center}

\end{minipage}};
\draw  [fill={rgb, 255:red, 244; green, 136; blue, 138 }  ,fill opacity=1 ]  (548,208) -- (617,208) -- (617,254) -- (548,254) -- cycle  ;
\draw (551,212) node [anchor=north west][inner sep=0.75pt]   [align=left] {\begin{minipage}[lt]{44.87pt}\setlength\topsep{0pt}
\begin{center}
Translate\\Solution
\end{center}

\end{minipage}};
\draw  [fill={rgb, 255:red, 244; green, 136; blue, 138 }  ,fill opacity=1 ]  (429, 231) circle [x radius= 55.15, y radius= 29.7]   ;
\draw (391,212) node [anchor=north west][inner sep=0.75pt]   [align=left] {\begin{minipage}[lt]{53.19pt}\setlength\topsep{0pt}
\begin{center}
Is Solution \\Valid?
\end{center}

\end{minipage}};
\draw  [fill={rgb, 255:red, 244; green, 136; blue, 138 }  ,fill opacity=1 ]  (252, 228.5) circle [x radius= 53.74, y radius= 44.55]   ;
\draw (215,199) node [anchor=north west][inner sep=0.75pt]   [align=left] {\begin{minipage}[lt]{51.86pt}\setlength\topsep{0pt}
\begin{center}
Is Solution\\Sufficiently\\Good?
\end{center}

\end{minipage}};
\draw  [fill={rgb, 255:red, 244; green, 136; blue, 138 }  ,fill opacity=1 ]  (32,208) -- (122,208) -- (122,254) -- (32,254) -- cycle  ;
\draw (35,212) node [anchor=north west][inner sep=0.75pt]   [align=left] {\begin{minipage}[lt]{58.85pt}\setlength\topsep{0pt}
\begin{center}
Postprocess\\Solution
\end{center}

\end{minipage}};
\draw  [fill={rgb, 255:red, 244; green, 136; blue, 138 }  ,fill opacity=1 ]  (48,118) -- (96,118) -- (96,143) -- (48,143) -- cycle  ;
\draw (51,122) node [anchor=north west][inner sep=0.75pt]   [align=left] {STOP};
\draw  [fill={rgb, 255:red, 250; green, 185; blue, 48 }  ,fill opacity=1 ]  (292,107) -- (392,107) -- (392,132) -- (292,132) -- cycle  ;
\draw (295,111) node [anchor=north west][inner sep=0.75pt]   [align=left] {\begin{minipage}[lt]{65.63pt}\setlength\topsep{0pt}
\begin{center}
Modify QUBO
\end{center}

\end{minipage}};
\draw (153,201) node [anchor=north west][inner sep=0.75pt]   [align=left] {Yes};
\draw (331,202) node [anchor=north west][inner sep=0.75pt]   [align=left] {Yes};
\draw (268,151) node [anchor=north west][inner sep=0.75pt]   [align=left] {No};
\draw (391,161) node [anchor=north west][inner sep=0.75pt]   [align=left] {No};

\end{tikzpicture}
\end{figure}


\subsection{Closed Undirected CPP}
\label{Undirected Explanation}
According to a literature review completed by the authors, the first and only previous work done on the subject of using a QA to solve the CPP in any form was done by Siloi et al. in ``Investigating the Chinese Postman Problem on a Quantum Annealer''~\cite{Siloi} which solved the Closed Undirected CPP.

Below we outline our implementation for solving the Closed Undirected CPP in Algorithm \ref{CUCPPAlgorithm}. The reason why this algorithm works is a consequence of the following two theorems \cite{graphtextbook}.

\begin{algorithm}
\caption{Solving the Closed Undirected CPP on a QA}
\label{CUCPPAlgorithm}
\begin{algorithmic}[1]

\Procedure{Routing}{$G$}        \Comment{$G$ is an undirected graph}
    \State Find all nodes of odd degree in $G$
    \State Create QUBO for $G$, QUBO($G$)  
    \State Run Quantum Annealing Process on QUBO($G$) \Comment{Do $N$ times}
    \State Identify lowest energy solution, $S$
    \State Intepret $S$ as perfect pairing amongst odd degree nodes of $G$
    \If{$S$ is not a valid solution}
        \State Modify QUBO
        \State GOTO Line 4
    \EndIf
    \State Create new graph $G'$ by adding perfect pairing edges to $G$
    \State Find Eulerian Circuit in $G'$, $E'$
    \State Replace added edges in $E'$ with corresponding path to produce path $E$
\EndProcedure

\end{algorithmic}
\end{algorithm}


\begin{theorem}
(Fundamental Theorem of Graph Theory) Given an undirected graph, $G$, the sum of the degrees of every vertex in $G$ is even and is twice the number of edges. 
\end{theorem}

\begin{theorem}
(Euler Circuit Criterion) Given an undirected graph, $G$, $G$ contains an Eulerian circuit if and only if every vertex in $G$ is of even degree. 
\end{theorem}

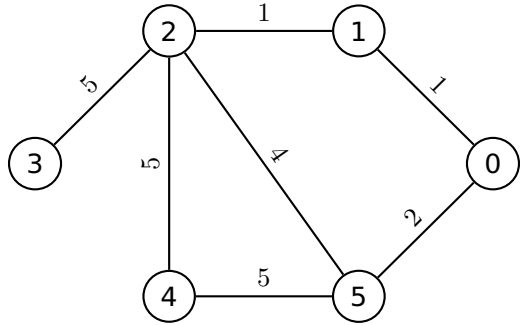
\begin{figure}
    \centering
    \caption{Undirected CPP algorithm example. The first graph introduces an example of an undirected
    weighted graph. Below it is a solution to the Undirected CPP for that graph presented in vertex order notation. The way to read this solution is to start at the first vertex in the list and then traverse the edge which connects to the next vertex in the list. One should continue this process until one reaches the end of the list. The second graph is the same as the first, except one extra edge was added with the weight of the shortest path between those two vertices. That edge was added because it makes every vertex have even degree and has the smallest walk weight between the two vertices.}
    \label{SiloiExample}
\begin{tikzpicture}[node distance={25mm}, thick, main/.style = {draw, circle}][h!] 
\node[main] (1) {$1$}; 
\node[main] (2) [left of=1] {$2$}; 
\node[main] (3) [below left of=2] {$3$}; 
\node[main] (4) [below right of=3] {$4$}; 
\node[main] (5) [right of=4] {$5$}; 
\node[main] (6) [above right of=5] {$0$}; 
\draw (3) -- node[above, sloped, pos=0.5] {5} (2); 
\draw (2) -- node[above, sloped, pos=0.5] {1} (1); 
\draw (1) -- node[above, sloped, pos=0.5] {1} (6); 
\draw (6) -- node[above, sloped, pos=0.5] {2} (5); 
\draw (5) -- node[above, sloped, pos=0.5] {5} (4);
\draw (4) -- node[above, sloped, pos=0.5] {5} (2);
\draw (5) -- node[above, sloped, pos=0.5] {4} (2);
\end{tikzpicture} 
$$\text{solution: } [2,4,\textcolor{blue}{5,2,3},2,5,0,1,2]$$
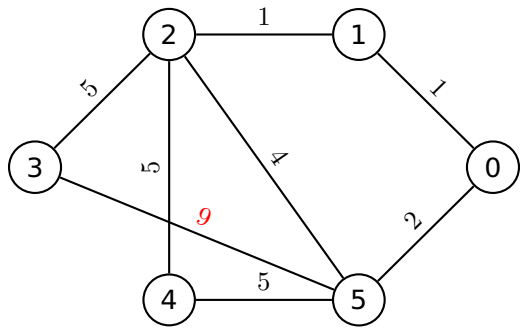
\begin{tikzpicture}[node distance={25mm}, thick, main/.style = {draw, circle}][h!] 
\node[main] (1) {$1$}; 
\node[main] (2) [left of=1] {$2$}; 
\node[main] (3) [below left of=2] {$3$}; 
\node[main] (4) [below right of=3] {$4$}; 
\node[main] (5) [right of=4] {$5$}; 
\node[main] (6) [above right of=5] {$0$}; 
\draw (3) -- node[above, sloped, pos=0.5] {5} (2); 
\draw (2) -- node[above, sloped, pos=0.5] {1} (1); 
\draw (1) -- node[above, sloped, pos=0.5] {1} (6); 
\draw (6) -- node[above, sloped, pos=0.5] {2} (5); 
\draw (5) -- node[above, sloped, pos=0.5] {5} (4);
\draw (4) -- node[above, sloped, pos=0.5] {5} (2);
\draw (5) -- node[above, sloped, pos=0.5] {4} (2);
\draw (3) -- node[above, sloped, pos=0.5, color = red] {9} (5);
\end{tikzpicture} 
$$\text{solution: } [2,4,\textcolor{red}{5,3},2,5,0,1,2]$$
\end{figure}

Upon inspection it should be obvious that if an Eulerian circuit exists in your graph, then, that Eulerian circuit will be an optimal solution. Additionally, if there exists an Eulerian circuit, then, it can be found in polynomial time \cite{EulerCurcuit}. One good implementation of the algorithm for finding an Eulerian Circuit is in \emph{NetworkX} \cite{networkx}. By the Fundamental Theorem of Graph Theory we conclude that there must be an even number of vertices in $G$ of odd degree. For any closed walk in $G$, considering the walk as its own graph (possibly multi-graph), we know the in-degree must equal the out-degree for each vertex in the walk, and hence in $G$. This leads us to realize that for every vertex of odd degree, we must reuse one of its edges in any Undirected Closed CPP solution. 

Let us say we have a solution to the Undirected Closed CPP. Let us then say that we are at the 
first point in our walk where we exit a vertex of odd degree.  As there are an even number of 
vertices of odd degree in $G$, we know there must be at least one other vertex of odd degree. As 
our walk traverses every edge in $G$ we know we must at some point enter another vertex of odd 
degree and take the first such instance after exiting our previous odd degree node. One 
could replace the path taken between those two vertices with an edge whose weight is equal to that
of the path's walk weight. Then in this modified graph, the two vertices whose degree was odd 
would now be even. We then repeat this process until all the vertices have even degree. Now we 
conclude that finding a solution to the Undirected Closed CPP is reduced to finding a perfect 
pairing of the vertices of odd degree which adds the least amount 
of weight. It is this problem that we formulate as a QUBO and solve using a QA.

Let $G$ be an undirected graph with vertices of odd degree, $\{v_1,...,v_d\}$. Let $x_{i,j}$ be a binary variable for $i,j\in\{1,...,d\}$ with $i<j$. Note we will use $x_{i,j}$ and $x_{j,i}$ to represent the same binary variable. The variable $x_{i,j}$ 
with value one or zero represents, respectively, vertices $v_i, v_j$ being paired together or not paired together. The symbol $W_{i,j}$ will be a constant representing the weight of the shortest path between $v_i$ and $v_j$. The symbol $P$ will be a positive constant 
whose use will be described shortly. The QUBO problem is then defined as: $$min\{(\sum\limits_{i = 1}^{d-1}\sum\limits_{j = i+1}^d W_{i,j}x_{i,j}) + P(\sum\limits_{i=1}^{d}(1-\sum\limits_{\substack{j = 1 \\ j\neq i}}^dx_{i,j})^2)\}.$$ To understand this QUBO, let us 
examine it term by term. First consider: $$\sum\limits_{i=1}^{d}(1-\sum\limits_{\substack{j = 1 \\ j\neq i}}^dx_{i,j})^2.$$ We shall 
label this part of the equation as $C$, for constraint. This part of the equation is to make sure a perfect pairing is formed. If $C$ 
equals zero, we may interpret this as: ``for every i, the vertex $v_i$ is paired with exactly one vertex $v_j$ with $j\neq i$.'' Now we
consider: $$min\{(\sum\limits_{i = 1}^{d-1}\sum\limits_{j = i+1}^d W_{i,j}x_{i,j}).$$ We shall label this part of the equation $M$ for minimize. Let us assume that $C$ = 0 and thus we have a perfect pairing. Then the minimum value of $M$, and hence the whole QUBO, will 
correspond to the perfect pairing amongst the odd degree vertices which adds the least total weight. In other words, the outputs of our
binary variables, $x_{i,j}$, which correspond to the minimal value of the QUBO, will in turn correspond to a solution of the Closed 
Undirected CPP. \textit{Note also the number of variables in this QUBO is easy to compute as it is just ${d \choose 2}$, where once again, $d$ is the number of vertices of odd degree; also, the variables in the QUBO are not fully connected}. Now we shall talk about the use and importance of the constant $P$. As we get a perfect pairing if 
and only if $C = 0$, we must make sure that the QUBO will be, in net, penalized for breaking this condition. Consider for a moment a new graph with two vertices and one edge between them with $W_{1,2} > 2$, and that 
we set $P = 1$. If we set $x_{1,2} = 0$, we won't get a perfect 
pairing, but we will minimize the QUBO overall.  This is because 
relatively $C$ will increase by $2$ while $M$ decreases by more than $2$ by switching $x_{1,2} = 1$ to $x_{1,2} = 0$. Thus we see the need for a constant $P$ since it will increase 
the cost incurred by breaking the condition $C = 0$. In theory, $P$ should be set equal to some arbitrarily large number as we require 
that $C = 0$. This does not work in practice because when setting up for the annealing process, all binary variable constants are 
scaled to fit into an interval. Due to a limit on the sensitivity of the method, values sufficiently close to zero will be treated 
as zero. In practice the value that should be chosen for $P$ will depend on the graph itself and should be large enough to force $C = 0$, but small enough to not overshadow the $M$ component.

As a quick example, the QUBO one would get from the graph in Figure \ref{SiloiExample} would be simply, $$9x_{3,5} + P(1-x_{3,5})^2,$$ which if we let $P = 10$, for example, could be written as $$10(x_{3,5})^2 - 11x_{3,5} + 10.$$ Upon inspection one can see that the value of $x_{3,5}$ which minimizes this equation is $x_{3,5} = 1$ for a total value of $9$ as expected.

\subsection{A General Approach to the Chinese Postman Problem}\label{BGC}

We now outline an algorithm for solving a more general class of CPPs on a QA. We show how to solve variants \ref{undirected variant} through \ref{windy variant} from section \ref{variants} and discuss results for those variants in section \ref{results}. Algorithm \ref{GeneralAlgorithm} is an outline of the general algorithm.

\begin{algorithm}[h!]
\caption{Quantum Annealing for a CPP}
\label{GeneralAlgorithm}
\begin{algorithmic}[1]

\Procedure{Routing}{$G$}
    \State Collect Graph Data
    \State Choose maximum length of walk
    \State Create QUBO for $G$, QUBO($G$)
    \State Run Quantum Annealing Process on QUBO($G$) \Comment{Do 10-10000 times}
    \State Identify lowest energy solution, $S$
    \State Intepret $S$ as walk in $G$
    \If{$S$ is not a valid solution}
        \State Modify QUBO
        \State GOTO Line 5
    \EndIf
\EndProcedure

\end{algorithmic}
\end{algorithm}

Now we shall explain the construction of the QUBO. Let $G$ be a graph, directed, undirected or 
mixed. Let $V$ be the set of vertices in $G$. Let $i_{max}\in\mathbb{N}$. The constant $i_{max}$ 
represents the maximum number of edges we will allow to be traversed in the walk and we will 
discuss how to choose $i_{max}$ later.  Our binary variables will be ${}_ie_{j,k}$ and 
${}_{2^r}s_{j,k}$ for $j,k\in V$, such that there is an edge going from vertex $j$ to vertex $k$, 
$i\in\{0,...,i_{max}\}$, and ${r\in\{0,...,ceiling(\log(i_{max}))\}}$. Let $W_{i,j}$ be the weight of
the edge going from vertex $i$ to vertex $j$. The mental picture we will use to guide our thinking
is that of choosing the steps in our path in an ordered manner. The variable ${}_ie_{j,k}$ 
taking the value of one in the solution to the QUBO will correspond to an
instruction to traverse the edge going from vertex $j$ to vertex $k$ in the i${}^{\text{th}}$ 
step of the walk. The variables, ${}_{2^r}s_{j,k}$, are slack variables, which are helpful in 
setting up some of our inequality conditions. Recall that we are presetting the number of 
steps we will take in the walk in our graph. As we do not know a priori how many steps we need for our walk, let us assume we overestimate. To compensate for this overestimation, we will allow 
repetition in the steps we take in our walk so as to allow a shorter walk than the one 
predetermined in our set-up. For example, ${}_ie_{j,k}$ and ${}_{i+1}e_{j,k}$ will be allowed to 
simultaneously evaluate to 1. 

Using this set-up, let us talk about what conditions and constraints need to be met to create a legal path. First, at any given step in our walk, the walk should traverse precisely one edge. This can be phrased as: for all $i$, there is a unique pair, $(j,k)$, such that ${}_ie_{j,k} = 1$. Written as a constraint for a QUBO this is,
$$C_{\text{one\_edge}} = \sum\limits_{i}(1-\sum\limits_{j,k}{}_ie_{j,k})^2.$$
The next constraint is that we do not want our walk to 'jump' around in our graph, we only want graph walks. To do this we will require that if ${}_ie_{j,k}$ and ${}_{i+1}e_{r,s}$ are both one, then either $k=r$ and the walk is legal at this location or $j=r$ and $k=s$ and we have a repetition edge whose necessity for existence was explained above. Written as a constraint for a QUBO this is,
$$C_{\text{adjacency}} = \sum\limits_i\sum\limits_{j,k}\sum\limits_{\substack{r,s \\ r\neq k \\ \text{and} \\ r \neq j \text{ or } s \neq k}}{}_ie_{j,k}\cdot {}_{(i+1)}e_{r,s}.$$
The next constraint is that we want to make sure that every edge in the graph which is required to be included into the walk is included. This will be phrased as: for any directed edge from vertex $j$ to vertex $k$ which should be included in the walk, there exists at least one $i$ such that ${}_ie_{j,k} = 1$. And for any undirected edge between vertex $j$ and vertex $k$, there exists at least one $i$ such that ${}_ie_{j,k} = 1$ or ${}_ie_{k,j} = 1$. It is for this kind of inequality that we need the slack variables. Written as a constraint for a QUBO this constraint is:
\begin{align*}
C_{\text{required\_directed}} =& \sum\limits_{(j,k)}((1-\sum\limits_i ({}_ie_{j,k})) + \sum\limits_r 2^r{}_{2^r}s_{j,k})^2,\\
C_{\text{required\_undirected}} =& \sum\limits_{[j,k]}((1-\sum\limits_i ({}_ie_{j,k} + {}_ie_{k,j})) + \sum\limits_r 2^r{}_{2^r}s_{j,k})^2, \text{and}\\
C_{\text{required}} =& ~C_{\text{required\_directed}} + C_{\text{required\_undirected}}.
\end{align*}

The next constraint which may occur is a required start and/or stop location. Unlike the constraints above which increase the required connectivity between variables in the QUBO, which makes it more difficult to embed on the hardware, we can use this constraint to decrease the number of variables needed and also decrease the connectivity between variables in the QUBO. We can do this by not creating unneeded variables, which means we would also omit those variables from the constraints above. If a start location is required then we shall not include variables ${}_ie_{j,k}$ for which the edge $(j,k)$ cannot be reached in $i$ steps, accounting for the type of repetition we spoke of previously. This is done similarly at the end of the walk if an end location is specified.  Let us consider as an example the first graph in Figure \ref{SiloiExample} and let us specify the starting vertex to be vertex $3$. Then for the variables, ${}_ie_{j,k}$, for which $i = 0$, the first step, we only have ${}_0e_{3,2}$. For $i = 1$ we only have ${}_1e_{3,2}$, ${}_1e_{2,1}$, ${}_1e_{2,3}$, ${}_1e_{2,4}$, and ${}_1e_{2,5}$. In this way we forcibly achieve any start or stop constraint. 

Let us take a brief pause from talking about constraints to introduce the part of the QUBO which will lead us to picking the best walk amongst all the legal walks which meet our requirements. The constraints above will all be zero if the walk chosen is legal and meets the preset requirements. This following part though, apart from some trivial cases, will be non-zero and is the `meat' of what we are trying to minimize. The essence of this part of the QUBO is that we want to add up all the weights of all the edges we traverse while not double counting weights when an edge is repeated. This is done as follows,
$$M = \sum\limits_{i>0}\sum\limits_{j,k}W_{j,k}({}_ie_{j,k}\cdot (1 - {}_{(i-1)}e_{j,k})) + \sum\limits_{j,k}W_{j,k}({}_0e_{j,k}).$$
Before we proceed further, it is useful to note that we have produced enough constraints to handle all combinations of variants \ref{undirected variant} through \ref{windy variant} for the CPP using the QUBO,
$$Q = M + P_{\text{required}}\cdot C_{\text{required}} + P_{\text{adjacency}}\cdot C_{\text{adjacency}} + P_{\text{one\_edge}}\cdot C_{\text{one\_edge}},$$
where the $'P'$ variable's are positive real numbers which are used to scale the weight of the constraints. How to choose those values will be discussed later. 

There is a alternative optimization to the start/stop optimization which will now be described in brief. The modifications to the QUBO related to this change are similar to those at the end of section \ref{even more general}. Rather than using the edge repetition method used to handle $i_{\text{max}}$ over-estimations, one can instead use a terminal vertex method. The idea of the terminal vertex method is to add one additional vertex, the terminal vertex, to the graph which will represent the end of the walk. One then needs to add the appropriate edges. If a variable corresponding to an edge leading to the terminal vertex at step $i$ in the walk takes the value $1$ in the QUBO solution, this is interpreted as the walk ending at step $i$. The edges one should add are a directed edge from any vertex the walk is allowed to end at to the terminal vertex and a directed edge from the terminal vertex to itself.
The terminal vertex edges should only be included in the above QUBO at time steps greater than or equal to the cardinality of the set of required edges. The terminal vertex method allows us to do two things. One, we may remove the edge repetition from $M$, decreasing the connectivity between variables in the QUBO, and two, we may remove the edge repetition from determining which edges are possible to reach in the $i^{\text{th}}$ step in the start/stop optimization above, which decreases the number of variables needed. However, one must account for the increase of variables and variable connectivity induced by including the terminal vertex itself. The two optimizations both do better on different graphs, depending on the graph topology. The authors have implemented both methods and choose for each problem whichever uses the least variables. Recall that the above QUBO was implemented 
with results shown in Section \ref{results}.

\subsection{Expanding the General Approach to the CPP}
\label{even more general}

The following variants, variants \ref{k variant} through \ref{hierachy variant}, have not yet been implemented on a quantum annealing device. We provide QUBO equations for implementing these variants and discuss why the equations are valid.

To include variant \ref{turning variant} we need to only add one additional constraint. One may recall that variant \ref{turning variant} includes additional information in the form of 3-tuples, (edge-in, edge-out, bonus weight). Similar to the ${}_ie_{j,k}$ variables, let us write the 3-tuple as $((j,k),(k,r),x_{j,k,r})$ where $j,k,r\in V$ such that $(j,k), (k,r)$ are edges in the graph. Then the QUBO constraint can be written as,
$$C_{\text{turn}} = \sum\limits_{i,j,k,r}x_{j,k,r}({}_ie_{j,k}\cdot{}_{(i+1)}e_{k,r}),$$
which in conjunction to what we had before would create the QUBO,
$$Q + P_{\text{turn}}\cdot C_{\text{turn}}.$$
Observe the turning constraint does not add any more variables, but does increase the connectivity between variables in the QUBO. 

Adding variants \ref{k variant} and \ref{service variant} requires additional modifications to the QUBO construction above. Let us start with variant \ref{service variant}, service-based traversal. We shall replace the variables ${}_ie_{j,k}$ with ${}_ie^s_{j,k}$, ${}_ie^t_{j,k}$. The ${}_ie^s_{j,k}$ variable equaling 1 corresponds to servicing the edge going from vertex $j$ to vertex $k$ on the $i^{\text{th}}$ step of the walk. Setting ${}_ie^t_{j,k}$ = 1 will correspond to merely traversing the edge going from vertex $j$ to vertex $k$ on the $i^{\text{th}}$ step of the walk. We will also not use the ${}_{2^r}s_{j,k}$ variables. The modifications to the constraints are as follows:
\begin{align*}
C_{\text{one\_edge}} =& \sum\limits_{i}(1-\sum\limits_{j,k}({}_ie^s_{j,k} + {}_ie^t_{j,k}))^2,\\
C_{\text{adjacency}} =& \sum\limits_i\sum\limits_{j,k}(\sum\limits_{\substack{r,s \\ r\neq k \\ \text{and} \\ r \neq j \text{ or } s \neq k}}({}_ie^t_{j,k}\cdot {}_{(i+1)}e^t_{r,s})\\ &+ \sum\limits_{\substack{r,s \\ r\neq k}}({}_ie^s_{j,k}\cdot {}_{(i+1)}e^t_{r,s} + {}_ie^s_{j,k}\cdot {}_{(i+1)}e^s_{r,s} + {}_ie^t_{j,k}\cdot {}_{(i+1)}e^s_{r,s})),\\
C_{\text{required\_directed}} =& \sum\limits_{(j,k)}(1-\sum\limits_i {}_ie^s_{j,k})^2,\\
C_{\text{required\_undirected}} =& \sum\limits_{[j,k]}(1-\sum\limits_i ({}_ie^s_{j,k} + {}_ie^s_{k,j}))^2,\\
C_{\text{required}} =& C_{\text{required\_directed}} + C_{\text{required\_undirected}}, \text{and}\\
C_{\text{turn}} =& \sum\limits_{i,j,k,r}x_{j,k,r}({}_ie^s_{j,k}\cdot{}_{(i+1)}e^s_{k,r} + {}_ie^s_{j,k}\cdot{}_{(i+1)}e^t_{k,r} \\&+ {}_ie^t_{j,k}\cdot{}_{(i+1)}e^s_{k,r} + {}_ie^t_{j,k}\cdot
{}_{(i+1)}e^t_{k,r}).
\end{align*}
Let $W^s_{j,k}$ be the weight corresponding to servicing the edge going from vertex $j$ to vertex $k$, while $W^t_{j,k}$ is the weight corresponding to just traversing that edge. Then,
$$M = \sum\limits_{i>0}\sum\limits_{j,k}(W^t_{j,k}\cdot {}_ie^t_{j,k}\cdot (1 - {}_{(i-1)}e^t_{j,k}) + W^s_{j,k}\cdot {}_ie^s_{j,k}) +  \sum\limits_{j,k}(W^s_{j,k}\cdot{}_0e^s_{j,k} + W^t_{j,k}\cdot{}_0e^t_{j,k}).$$
One caveat to the service traversal variant QUBO set-up is that if the solution involves only service steps and no traversal steps, we won't be able to get the optimal solution if we don't a priori know the precise number of steps our walks need as we only allow repetition on traversal steps. This is easily avoided in practice however as the only way a walk will only use service steps is if there is an Eulerian circuit on the subset of edges required, which as stated before, is computationally easy to determine. One benefit is that it is easy to to make the service based traversal hierarchical, like variant \ref{hierachy variant}, for any partially ordered set (i.e. some edges must be serviced prior to others) without introducing more variables. Solving variant \ref{hierachy variant} is achieved by adding on the following constraint. Let $x_{(j,k),(r,s)} = 1$ if the edge going from vertex $j$ to vertex $k$ must be serviced prior to the edge going from vertex $r$ to vertex $s$, and $0$ otherwise. Note $x_{(j,k),(r,s)}$ is a given value in the problem and not a variable the annealing device solves for. Let the other notation be similar. Recall $(*,*)$ is for directed edges while $[*,*]$ is for undirected edges. Then,
\begin{align*}
C_{\text{hierarchy}} =& \sum\limits_{i_0}\sum\limits_{i_1<i_0,j,k,r,s}(x_{(j,k),(r,s)}\cdot{}_{i_0}e^s_{j,k}\cdot{}_{i_1}e^s_{r,s})\\
&+ (x_{[j,k],(r,s)}\cdot({}_{i_0}e^s_{j,k}\cdot{}_{i_1}e^s_{r,s} + \cdot{}_{i_0}e^s_{k,j}\cdot{}_{i_1}e^s_{r,s}))\\
&+ (x_{(j,k),[r,s]}\cdot({}_{i_0}e^s_{j,k}\cdot{}_{i_1}e^s_{r,s} + \cdot{}_{i_0}e^s_{j,k}\cdot{}_{i_1}e^s_{s,r}))\\
&+ (x_{[j,k],[r,s]}\cdot({}_{i_0}e^s_{j,k}\cdot{}_{i_1}e^s_{r,s} + \cdot{}_{i_0}e^s_{j,k}\cdot{}_{i_1}e^s_{s,r}\\ 
&+ {}_{i_0}e^s_{k,j}\cdot{}_{i_1}e^s_{s,r} + \cdot{}_{i_0}e^s_{k,j}\cdot{}_{i_1}e^s_{s,r})).
\end{align*}

The final variant to talk about is variant \ref{k variant}, the $k$-Postman Problem. As we already use $k$, we shall suppose there are $l$ postmen. This variant provides us with the opportunity to introduce a slightly different idea. For variant \ref{k variant} we will need to use a slight modfication of the terminal vertex paradigm. We shall use the binary variables ${}_ie^a_{j,k}$, ${}_{2^r}s^a_{j,k}$, and ${}_irest^{a}$ for $a\in\{1,...,l\}$, all else the same. The $a$ index will refer to the $a^{\text{th}}$ postman walk. So ${}_ie^a_{j,k} = 1$ will correspond to the $a^{\text{th}}$ postman traversing the edge going from vertex $j$ to vertex $k$ on the $i^{\text{th}}$ step of their walk. The ${}_{2^r}s^a_{j,k}$ binary variable will be used as a slack variable similarly to before. When the variable ${}_irest^{a}$ equals one, this will correspond to the $a^{\text{th}}$ postman at their walk's endpoint on the $i^{\text{th}}$ step and is their terminal vertex from the terminal vertex method described in section \ref{BGC}. Let $W^a_{j,k}$ represent the weight corresponding to the $a^{\text{th}}$ postman traversing the edge going from vertex $j$ to vertex $k$. The variations to the QUBO constraints are listed below:
\begin{align*}
C_{\text{one\_edge}} =& \sum\limits_{i,a}(1-\sum\limits_{j,k}({}_ie^a_{j,k}) - {}_irest^a)^2,\\
C_{\text{adjacency}} =& \sum\limits_{i,a}\sum\limits_{j,k}((\sum\limits_{\substack{r,s \\ r\neq k}}({}_ie^a_{j,k}\cdot {}_{(i+1)}e^a_{r,s})) + ({}_irest^a\cdot {}_{i+1}e^a_{j,k})),\\
C_{\text{required\_directed}} =& \sum\limits_{(j,k)}(1-\sum\limits_a(\sum\limits_i ({}_ie^a_{j,k}) - \sum\limits_r 2^r{}_{2^r}s^a_{j,k}))^2,\\
C_{\text{required\_undirected}} =& \sum\limits_{[j,k]}(1-\sum\limits_a(\sum\limits_i ({}_ie^a_{j,k} + {}_ie^a_{k,j}) - \sum\limits_r 2^r{}_{2^r}s^a_{j,k}))^2,\\
C_{\text{required}} =& C_{\text{required\_directed}} + C_{\text{required\_undirected}},\\
C_{\text{turn}} =& \sum\limits_{i,j,k,r,a}x_{j,k,r}({}_ie^a_{j,k}\cdot{}_{(i+1)}e^a_{k,r}), \text{and}\\
M =& \sum\limits_{a,i,j,k}W^a_{j,k}\cdot {}_ie^a_{j,k}.
\end{align*}

Consider a use case where no two postmen may occupy the same edge going in the same direction at the same time. An example of an advantage one gains for using this resting paradigm over the repeated edge paradigm is that it becomes easy to create a constraint to avoid such collisions, observe,

$$C_{\text{collisions}} = \sum\limits_{i,a,b,j,k}{}_ie^a_{j,k}\cdot {}_ie^b_{j,k}.$$

Similar constraints can be made for avoiding collisions going in the opposite direction along edges and for avoiding collisions at vertices. Additionally, if $W^a_{j,k}\in\mathbb{N}_0$ for all $i,j,a$, then one may consider the $k$-Postman Problem With Capacity. Suppose one is given $\{c_1,...,c_l\}\subset\mathbb{N}$, where the $a^{\text{th}}$ postman is limited to walks of weight less than $c_a$. Then, by introducing the slack variables ${}_{2^y}slack^a$ for $y\in\{0,...,ceiling(\log(c_a))\}$ for $a\in\{1,...,l\}$, we have the constraint,
$$C_{\text{capacitance}} = \sum\limits_{a}(c_a - \sum\limits_{i,j,k}(W^a_{j,k}{}_ie^a_{j,k}) - \sum\limits_y(2^y{}_{2^y}slack^a))^2.$$ The slack variables in the equation allow the postmen to have walks with walk weights less than their maximum capacity without punishing such solutions.

One may also note that it is possible to combine variant \ref{k variant} and \ref{service variant} together with a little thought, however we shall abstain from doing so to save ourselves from the additional complex
notation it would create.

\section{Results}\label{results}

In this section, we will consider the results of two kinds of experiments for both the Closed Undirected CPP and the general CPP. The first kind is a parameter study for various tuneable parameters for running the problem directly on quantum hardware. The second kind is a comparison of results between various purely quantum, classical, and quantum-classical solutions to the same CPP problem. A reminder to help clarify the results shown is that 
the goal is to minimize the solution so as to minimize the weight of the walk. 

\subsection{Closed Undirected CPP Parameter Study}

Let us begin with a parameter study of the Closed Undirected CPP. The parameters tuned were the constraint weight, the sample number, the intersample correlation, the number of spin reversal transforms, and the annealing time.

The constraint weight is the $P$ variable discussed in section \ref{Undirected Explanation}. The constraint weight needs to be large enough to encourage valid solutions to the Closed Undirected CPP, but small enough to not overshadow the rest of the QUBO.

\begin{table}[h]
    \centering
    \caption{$P$ Value Efficacy Directly on 2000Q}
    \label{CUCPP P Table}
    \begin{tabular}{*5c}
        \toprule
        & \multicolumn{4}{c}{$P$ Values} \\
        \cmidrule(lr){2-5}
        ~ & 10  & 30 & 50 & 70 \\    
        \midrule
        valid (\%) & 27.5  & 33.8 & 37.5 & 50\\
        time to solution (avg, s) & 3.11 & 3.05 & 2.89 & 3.11\\
        optimal (\%) & 23.8 & 25 & 26.3 & 26.3\\
        $\leq$ 10\% above optimum (\%) & 25 & 25 & 26.3 & 27.3 \\
        $\leq$ 25\% above optimum (\%) & 25 & 25 & 30 & 33.8 \\
        \bottomrule
    \end{tabular}
\end{table}
Some guidance is provided for understanding the tables.
In Table \ref{CUCPP P Table}, the number $3.05$ in column $30$ and row 'time to solution,' means that using a $P$ value of $30$, the average wall-clock time to get the solution across all such runs was $3.05$s. The wall-clock time is the time it took the algorithm to find a solution from a given QUBO. Using the Leap based methods this includes over the internet communication plus the time it took to validate the lowest energy solution. The number $30$ in column $50$ and row '$\leq$ 25\% above optimum' means that $30\%$ of the runs using a $P$ value of $50$ were valid and achieved a solution less than (meaning better than) or equal to $25\%$ above (above meaning worse than) the optimal solution. 

As one can see in Table \ref{CUCPP P Table}, there is a strong correlation between the percentage of valid solutions and the percentage of optimal solutions with the size of the $P$ value. In fact, this becomes more apparent when we separate the data by the size of the problem as in Table \ref{CUCPP P Table Split}. The constraint weight for the Closed Undirected CPP runs 
when done strictly on quantum hardware will be set to $70$. Next, we will consider the effect the sample number has on the result. The sample number is the number of times states are read from the quantum hardware.

\begin{table}[h]
    \centering
    \caption{$P$ Value Efficacy by Number of Odd Vertices Directly on 2000Q}
    \label{CUCPP P Table Split}
    \begin{tabular}{*5c}
        \toprule
        & \multicolumn{4}{c}{4 Odd vertices: $P$ Values} \\
        \cmidrule(lr){2-5}
        ~ & 10  & 30 & 50 & 70 \\    
        \midrule
        valid (\%) & 100  & 100 & 100 & 100\\
        optimal (\%) & 95 & 100 & 95 & 100\\
        $\leq$ 25\% above optimum (\%) & 95 & 100 & 95 & 100 \\
        \toprule
                & \multicolumn{4}{c}{6 Odd vertices: $P$ Values} \\
        \cmidrule(lr){2-5}
        ~ & 10  & 30 & 50 & 70 \\    
        \midrule
        valid (\%) & 10  & 30 & 50 & 85\\
        optimal (\%) & 0 & 0 & 10 & 5\\
        $\leq$ 25\% above optimum (\%) & 5 & 0 & 25 & 30 \\
        \toprule
                & \multicolumn{4}{c}{8 Odd vertices: $P$ Values} \\
        \cmidrule(lr){2-5}
        ~ & 10  & 30 & 50 & 70 \\    
        \midrule
        valid (\%) & 0  & 5 & 0 & 15\\
        optimal (\%) & 0 & 0 & 0 & 0\\
        $\leq$ 25\% above optimum (\%) & 0 & 0 & 0 & 5 \\
        \toprule
                & \multicolumn{4}{c}{10 Odd vertices: $P$ Values} \\
        \cmidrule(lr){2-5}
        ~ & 10  & 30 & 50 & 70 \\    
        \midrule
        valid (\%) & 0  & 0 & 0 & 0\\
        optimal (\%) & 0 & 0 & 0 & 0\\
        $\leq$ 25\% above optimum (\%) & 0 & 0 & 0 & 0 \\
        \bottomrule
    \end{tabular}
\end{table}

\begin{table}[h]
    \centering
    \caption{Sample Number Efficacy Directly on 2000Q}
    \label{CUCPP Sample Table}
    \begin{tabular}{*6c}
        \toprule
        & \multicolumn{4}{c}{Sample Numbers} \\
        \cmidrule(lr){2-6}
        ~ & 10  & 50 & 100 & 500 & 1000 \\    
        \midrule
        valid (\%) & 27.9  & 35.3 & 51.5 & 58.8 & 61.8\\
        time to solution (avg, s) & 2.85 & 2.87 & 3.06 & 2.76 & 3.08\\
        optimal (\%) & 19.1 & 23.5 & 32.4 & 39.7 & 47.1\\
        $\leq$ 10\% above optimum (\%) & 19.1 & 26.5 & 32.4 & 44.1 & 48.5 \\
        $\leq$ 25\% above optimum (\%) & 22.1 & 26.5 & 36.8 & 48.5 & 50 \\
        \bottomrule
    \end{tabular}
\end{table}

 A strong positive relationship between the validity and quality of the solutions with the number of samples is seen in Table \ref{CUCPP Sample Table}. Even when viewed by problem size, the monotone increasing relationship between validity and quality of solution with sample number is preserved across every problem size. This is true except for one instance, a problem of size 8 odd degree vertices where one solution of moderate quality ($\leq$ $25$\% above optimum) was found for the $500$ sample number, but not the $1000$ sample number. A sample number of $1000$ will be used for the Closed Undirected CPP when there is a choice on quantum hardware.
 
 \begin{table}[h]
    \centering
    \caption{Reduced Intersample Correlation Efficacy Directly on 2000Q}
    \label{CUCPP Intersample Table}
    \begin{tabular}{*3c}
        \toprule
        & \multicolumn{2}{c}{Intersample Correlation} \\
        \cmidrule(lr){2-3}
        ~ & Not Reduced  & Reduced \\    
        \midrule
        valid (\%) & 57.5  & 63.8 \\
        time to solution (avg, s) & 3.03 & 3.46 \\
        optimal (\%) & 43.8 & 40 \\
        $\leq$ 10\% above optimum (\%) & 46.3 & 43.8 \\
        $\leq$ 25\% above optimum (\%) & 48.8 & 47.5 \\
        \bottomrule
    \end{tabular}
\end{table}

 Next, we will explore the relation that reducing intersample correlation has with solutions in Table \ref{CUCPP Intersample Table}. The relationship between reducing intersample correlation and solution validity/quality is mixed with reducing intersample correlation slightly increasing validity of the solution and slightly decreasing solution quality. The time to solution is on average increased when reducing intersample correlation and so is not used moving forward when studying the Closed Undirected CPP.

\begin{table}[h]
    \centering
    \caption{Spin Reversal Transforms Efficacy Directly on 2000Q}
    \label{CUCPP Spin Table}
    \begin{tabular}{*5c}
        \toprule
        & \multicolumn{4}{c}{Number of Spin Reversal Tranforms} \\
        \cmidrule(lr){2-5}
        ~ & 0  & 10 & 30 & 100  \\    
        \midrule
        valid (\%) & 60  & 61.03 & 58.8 & 60\\
        time to solution (avg, s) & 3.31 & 3.36 & 3.91 & 5.04\\
        optimal (\%) & 36.3 & 43.8 & 40 & 42.5\\
        $\leq$ 10\% above optimum (\%) & 38.8 & 43.8 & 41.3 & 45 \\
        $\leq$ 25\% above optimum (\%) & 43.8 & 48.8 & 46.3 & 51.3 \\
        \bottomrule
    \end{tabular}
\end{table}

 We will continue our parameter study of the Closed Undirected CPP by looking at the effect the number of spin reversal transforms has on solutions in Table \ref{CUCPP Spin Table}. There is little relationship between the validity of the solution and the number of spin reversal transformations. The solution quality does improve when using spin reversal transformations, with the largest jump in quality coming from the first $10$ spin reversal transforms. The number of spin reversal transformations increases the time to solution significantly when a large number is used. Only $10$ spin reversal transformations will be used henceforth for the Closed Undirected CPP when on quantum hardware.
 
 \begin{table}[h]
    \centering
    \caption{Annealing Time Efficacy Directly on 2000Q}
    \label{CUCPP Annealing Table}
    \begin{tabular}{*6c}
        \toprule
        & \multicolumn{4}{c}{Annealing Time ($\mu$s)}\\
        \cmidrule(lr){2-6}
        ~ & 5  & 10 & 30 & 100 & 500 \\    
        \midrule
        valid (\%) & 55.9  & 58.8 & 60.3 & 61.8 & 72.1\\
        time to solution (avg, s) & 3.25 & 3.24 & 3.41 & 3.1 & 3.82\\
        optimal (\%) & 42.6 & 48.5 & 41.2 & 45.6 & 44.1\\
        $\leq$ 10\% above optimum (\%) & 44.1 & 50 & 42.6 & 45.6 & 44.1 \\
        $\leq$ 25\% above optimum (\%) & 47.1 & 51.5 & 52.9 & 50 & 48.5 \\
        \bottomrule
    \end{tabular}
\end{table}
 
 The final piece of our parameter study for the Closed Undirected CPP is to study the effect that annealing time has on solutions. There is a positive relationship between the validity of solutions and the annealing time apparent in Table \ref{CUCPP Annealing Table}. The relationship between the annealing time and the solution quality, however, are less clear. To find a middle ground, an annealing time of $100\mu $s was chosen moving forward when running the Closed Undirected CPP on quantum hardware.
 
 \subsection{Closed Undirected CPP Comparison Study}
 
 In this section we will compare the validity and quality of solutions between a brute force solution and various methods for finding solutions to the QUBO corresponding to the problem. 
 The brute force method only runs on problems with a
 maximum of $14$ vertices of odd degree, while the tabu algorithm (tabu) \cite{dwavedoc} has no limit on 
 size. Note that the tabu algorithm is a modified steepest descent algorithm 
 which keeps track of the locations of the best solutions found and 
 temporarily changes values in the QUBO to promote search diversity. Another method tested was the greedy 
 algorithm (greedy) \cite{dwavedoc} which has no limit on size and is just a steepest descent solver. Other methods 
 compared are the tabu algorithm both preprocessed
 \cite{greedytabu} and post-processed with a greedy algorithm (greedy tabu) 
 with no limit on size and simulated annealing (SA) with no limit on size. SA is a modified hill-climbing algorithm which improves solution  diversity, and often quality, over other hill-climbing algorithms by allowing worse solutions to be picked sometimes during the search \cite{SAref}.
 Quantum and quantum-hybrid methods compared include pure 
 2000Q (2000Q) with a maximum of $10$ vertices of odd degree, 2000Q 
 post-processed with a greedy algorithm (greedy 2000Q) with a maximum of $10$ 
 vertices of odd degree, \emph{qbsolv} on the 2000Q with no limit on size 
 (2000Q qbsolv), pure Advantage4.1 (Advantage) with a maximum of $18$ vertices of odd degree, Advantage4.1 post-processed with a greedy algorithm (greedy 
 Advantage) with a maximum of $18$ vertices of odd degree, and finally 
 \emph{qbsolv} on the Advantage4.1 (Advantage qbsolv) with no limit on size. 
 For some runs we also compared \emph{qbsolv} on both devices when using the 
 fixed embedding composite (f-2000Q qbsolv and f-Advantage qbsolv 
 respectively) versus the embedding composite in the \textit{ocean-dwave-sdk} \cite{dwavedoc}.
 
 
 When running on sufficiently small problems, solution quality is compared to the brute force method which finds the optimal answer. When on problems too large for brute force, we instead compare solutions with greedy tabu which usually only provides approximate answers.

 \begin{table}[h]
    \centering
    \caption{Comparison on graphs with $4$ or $6$ odd degree vertices}
    \label{CUCPP 4,6 Table}
    \begin{tabular}{*6c}
        \toprule
        & \multicolumn{5}{c}{Solution Quality: \% $\leq$ ---\% above optimum} \\
        \cmidrule(lr){2-6}
        ~ & 0  & 10 & 25 & 100 \\    
        \midrule
        \textbf{classical} & ~  & ~ & ~ & ~ \\    
        \midrule
        greedy & 27 & 33 & 58 & 95\\
        tabu & 100  & 100 & 100 & 100\\
        SA & 100 & 100 & 100 & 100\\
        \midrule
        \textbf{quantum} & ~  & ~ & ~ & ~ \\    
        \midrule
        2000Q & 90 & 93 & 97 & 100 \\
        Advantage & 77 & 83 & 92 & 100\\
        \midrule
        \textbf{hybrid} & ~  & ~ & ~ & ~ \\    
        \midrule
        greedy 2000Q & 100 & 100 & 100 & 100 \\
        greedy Advantage & 100 & 100 & 100 & 100\\
        2000Q qbsolv & 100  & 100 & 100 & 100\\
        Advantage qbsolv & 100 & 100 & 100 & 100 \\
        \bottomrule
    \end{tabular}
\end{table}

 We first compare these methods on some small problems of $4$ and $6$ odd degree vertices. Note an example of how Table \ref{CUCPP 4,6 Table} should be read is as follows: the number $33$ in the column labeled `$10$' and row labeled `greedy' means that $33\%$ of the greedy solutions were valid and at most $10\%$ above the optimal solution. One observes from Table \ref{CUCPP 4,6 Table} that on small problems we get perfect results in all methods except 2000Q, Advantage, and greedy. The 2000Q and Advantage still attained strong results on these small problems, while greedy only achieved mediocre results.
 
  \begin{table}[h]
    \centering
    \caption{Comparison on graphs with $8$ or $10$ odd degree vertices}
    \label{CUCPP 8,10 Table}
    \begin{tabular}{*6c}
        \toprule
        & \multicolumn{5}{c}{Solution Quality: \% $\leq$ ---\% above optimum} \\
        \cmidrule(lr){2-6}
        ~ & 0  & 10 & 25 & 100 \\    
        \midrule
        \textbf{classical} & ~  & ~ & ~ & ~ \\    
        \midrule
        greedy & 14 & 20 & 50 & 98\\
        tabu & 100  & 100 & 100 & 100\\
        SA & 100 & 100 & 100 & 100\\
        \midrule
        \textbf{quantum} & ~  & ~ & ~ & ~ \\    
        \midrule
        2000Q & 4.5 & 4.5 & 11 & 27 \\
        Advantage & 2.3 & 2.3 & 4.5 & 6.8\\
        \midrule
        \textbf{hybrid} & ~  & ~ & ~ & ~ \\    
        \midrule
        greedy 2000Q & 100 & 100 & 100 & 100 \\
        greedy Advantage & 100 & 100 & 100 & 100\\
        2000Q qbsolv & 100  & 100 & 100 & 100\\
        Advantage qbsolv & 100 & 100 & 100 & 100 \\
        \bottomrule
    \end{tabular}
\end{table}

Now in Table \ref{CUCPP 8,10 Table} we compare some slightly larger problems of graphs of $8$ and $10$ odd degree vertices respectively. 
In this larger problem we see that all of the methods get optimal results except 2000Q, Advantage, and greedy. However, this time, 2000Q, Advantage, and greedy got poor results, with 2000Q and Advantage drastically decreasing in efficacy. It is interesting to observe that despite the fact that greedy and 2000Q individually were ineffective, when used in concert optimal results were achieved. And this holds similarly for greedy and Advantage. The advantage of a greedy post-processing is not unique to this problem and was also used to improve solution quality in \cite{eigenvector} and \cite{qde}.

  \begin{table}[h]
    \centering
    \caption{Comparison on graphs with $16$ or $18$ odd degree vertices}
    \label{CUCPP 16,18 Table}
    \begin{tabular}{*7c}
        \toprule
        & \multicolumn{6}{c}{Solution Quality: \% $\leq$ ---\% above greedy tabu} \\
        \cmidrule(lr){2-7}
        ~ & -10 & -5 & 0  & 10 & 25 \\    
        \midrule
        \textbf{classical} & ~  & ~ & ~ & ~ \\    
        \midrule
        greedy & 0 & 0 & 3.3 & 3.3 & 17\\
        tabu & 0 & 10 & 70 & 97 & 100\\
        SA & 0 & 0 & 3.3 & 20 & 77 \\
        \midrule
        \textbf{quantum} & ~  & ~ & ~ & ~ \\    
        \midrule
        Advantage & 0 & 0 & 0 & 0 & 0\\
        \midrule
        \textbf{hybrid} & ~  & ~ & ~ & ~ \\    
        \midrule
        greedy Advantage & 3.3 & 6.7 & 57 & 83 & 100\\
        2000Q qbsolv & 3.3 & 17 & 100 & 100 & 100\\
        Advantage qbsolv & 3.3 & 23 & 97 & 100 & 100\\
        \bottomrule
    \end{tabular}
\end{table}

Table \ref{CUCPP 16,18 Table} shows results for problems which are too large
to be solved by brute force or to run directly on 2000Q hardware. In fact, 
these are the largest problems which can be run directly on the Advantage hardware. Once again, greedy and Advantage both performed poorly alone, but when used together produced results comparable to greedy tabu. The two \emph{qbsolv} methods almost always found solutions which were equal to or better than greedy tabu, with the better results occurring a non-negligible amount of the time.

  \begin{table}[h]
    \centering
    \caption{Comparison on graphs with $20$, $30$ and $50$ odd degree vertices}
    \label{CUCPP 20,30,50 Table}
    \begin{tabular}{*6c}
        \toprule
        & \multicolumn{5}{c}{Solution Quality: \% $\leq$ ---\% above greedy tabu} \\
        \cmidrule(lr){2-6}
        ~ & -5 & 0  & 10 & 25 \\    
        \midrule
        \textbf{classical} & ~  & ~ & ~ & ~ \\    
        \midrule
        greedy & 0 & 0 & 3.3 & 13\\
        tabu & 3.3 & 97 & 100 & 100\\
        SA & 0 & 0 & 0 & 6.7\\
        \midrule
        \textbf{hybrid} & ~  & ~ & ~ & ~ \\    
        \midrule
        2000Q qbsolv & 0 & 63 & 87 & 100\\
        f-2000Q qbsolv & 0 & 67 & 87 & 100\\
        Advantage qbsolv & 0 & 47 & 83 & 100\\
        f-Advantage qbsolv & 0 & 60 & 97 & 100\\
        \bottomrule
    \end{tabular}
\end{table}

In Table \ref{CUCPP 20,30,50 Table} we studied problems which were too large to run directly on quantum hardware. The fixed embedding produced better results than the non-fixed embeddings for \emph{qbsolv}, especially for the Advantage device. However, the fixed embedding took significantly more overhead time for finding the initial embedding and the fixed embedding \emph{qbsolv} used significantly more runs on the quantum hardware than the non-fixed embeddings did.

\subsection{General CPP Parameter Study}

Let us now examine our parameter study of the General CPP on the Advantage hardware with a greedy algorithm post-processing.

\begin{table}[h]
    \centering
    \caption{$P_{\text{one\_edge}}$ Value Efficacy on Advantage}
    \label{GCPP Single P Table}
    \begin{tabular}{*8c}
        \toprule
        & \multicolumn{7}{c}{$P_{\text{one\_edge}}$ Values} \\
        \cmidrule(lr){2-8}
        ~ & 30  & 40 & 50 & 60 & 70 & 80 & 90 \\    
        \midrule
        valid (\%) & 86.7  & 85.7 & 80 & 75 & 66.7 & 72.7 & 61.5\\
        time to solution (avg, s) & 40.8 & 40.8 & 26.3 & 31.7 & 26.221 & 32.002 & 30.928\\
        $\leq$ 0\% below SA (\%) & 66.7 & 66.7 & 60 & 55 & 58.3 & 72.7 & 46.2\\
        $\leq$ 10\% above optimum (\%) & 66.7 & 66.7 & 60 & 55 & 58.3 &72.7 & 46.2 \\
        $\leq$ 25\% above optimum (\%) & 73.3 & 71.4 & 64 & 60 & 66.7 & 72.7 & 53.8 \\
        \bottomrule
    \end{tabular}
\end{table}

In Table \ref{GCPP Single P Table} one can see that overall, there is a negative relationship between $P_{\text{one\_edge}}$ and solution validity and quality, with $P_{\text{one\_edge}} = 80$ being an exception. The parameter value chosen for problems run after this is $P_{\text{one\_edge}} = 40$. While doing strictly worse than $P_{\text{one\_edge}} = 30$ in terms of solution validity and quality overall, $P_{\text{one\_edge}} = 40$ actually did significantly better when looking at the larger end of problems able to run directly on the hardware in the 100-200 variable range.

\begin{table}[h]
    \centering
    \caption{$P_{adjacency}$ Value Efficacy on Advantage}
    \label{GCPP P Next Move Table}
    \begin{tabular}{*4c}
        \toprule
        & \multicolumn{3}{c}{$P_{adjacency}$ Values} \\
        \cmidrule(lr){2-4}
        ~ & 60  & 70 & 80 \\    
        \midrule
        valid (\%) & 77.3  & 81 & 90.3\\
        time to solution (avg, s) & 29.1 & 52.6 & 27.473\\
        $\leq$ 0\% below SA (\%) & 59.1 & 52.4 & 61.3\\
        $\leq$ 10\% above SA (\%) & 59.1 & 57.1 & 61.3 \\
        $\leq$ 25\% above SA (\%) & 63.6 & 57.1 & 67.7 \\
        \bottomrule
    \end{tabular}
\end{table}

Table \ref{GCPP P Next Move Table} shows a strong positive relationship between the validity of the solution and $P_{adjacency}$. If one separates the data out by problem size, $P_{adjacency} = 70$ does the best in terms of validity and solution quality for problems in the $0-100$ variable range, while $P_{adjacency} = 80$ does better for problems in the $100-250$ range. So 
the average of these two values is used moving forward setting $P_{adjacency} = 75$.

\begin{table}[h]
    \centering
    \caption{$P_{required}$ Value Efficacy on Advantage}
    \label{GCPP P required Table}
    \begin{tabular}{*5c}
        \toprule
        & \multicolumn{4}{c}{$P_{required}$ Values} \\
        \cmidrule(lr){2-5}
        ~ & 30 & 40 & 50 & 60 \\    
        \midrule
        valid (\%) & 96.4  & 85.3 & 82.9 & 85.3 \\
        time to solution (avg, s) & 32 & 29.4 & 47.5 & 33.3\\
        $\leq$ 10\% below SA (\%) & 0 & 5.88 & 0 & 2.94\\
        $\leq$ 0\% below SA (\%)  & 71.4 & 73.5 & 54.3 & 64.7\\
        $\leq$ 10\% above SA (\%) & 71.4 & 73.5 & 54.3 & 64.7 \\
        $\leq$ 25\% above SA (\%) & 82.1 & 76.5 & 62.9 & 64.7 \\
        \bottomrule
    \end{tabular}
\end{table}

As can be seen in Table \ref{GCPP P required Table} there is a generally negative trend for solution validity and solution quality with respect to $P_{required}$. When one looks at the data by problem size, $P_{required} = 30$ and $P_{required} = 40$ each do better on certain problem sizes.
So once again we take the average and set $P_{required} = 35$ moving forward.

\begin{table}[h]
    \centering
    \caption{Chain Strength Efficacy on Advantage}
    \label{GCPP chain strength Table}
    \begin{tabular}{*5c}
        \toprule
        & \multicolumn{4}{c}{Chain Strength} \\
        \cmidrule(lr){2-5}
        ~ & 400 & 500 & 800 & 900 \\    
        \midrule
        valid (\%) & 94.7 & 93.8 & 84.2 & 88.9 \\
        time to solution (avg, s) & 36.8 & 27.3 & 70.6 & 29\\
        $\leq$ 10\% below SA (\%) & 0 & 6.25 & 0 & 0\\
        $\leq$ 0\% below SA (\%)  & 68.4 & 56.3 & 52.6 & 55.6\\
        $\leq$ 10\% above SA (\%) & 68.4 & 62.5 & 57.9 & 55.6 \\
        $\leq$ 25\% above SA (\%) & 73.7 & 68.8 & 57.9 & 61.1 \\
        \bottomrule
    \end{tabular}
\end{table}

Table \ref{GCPP chain strength Table} highlights that having a very large chain strength can deteriorate both the validity and quality of solutions.
A chain strength of $400$ leads to the highest percentage of valid solutions and mostly the highest quality solutions. However, when one looks at the data broken up by problem size, one finds that a chain strength of $500$ gets generally better results on larger problems. So 
the chain strength moving forward has been set to $475$.

\subsection{General CPP Comparison Study}
\label{GCPP Comparison Study}

We compare results for the General CPP algorithm using tabu, greedy tabu, SA, Advantage, greedy Advantage, 2000Q qbsolv, and Advantage qbsolv. We compare how each method does on various problem sizes, for the sizes a method can run. The problem sizes are broken up into three categories: small as $0-250$ variables, medium as $250-1000$ variables, and large as $1000-3200$ variables. As a reminder, how these variables are formulated is explained in Section \ref{BGC}. Roughly speaking, small problems correspond to graphs with $3-4$ vertices, $25\%-75\%$ edge saturation, and all kinds of start/stop conditions. Medium problems correspond to graphs with $5-6$ vertices, and similar other data. Large problems were run on graphs with $9-10$ vertices, $25\%-50\%$ edge saturation, and similar other data.  We compare against SA for the small and medium sized problems. However, for the medium sized problems SA took a long time to compute. For the large problems, SA became prohibitively expensive to run. The large problems are compared against greedy tabu.

  \begin{table}
    \centering
    \caption{Comparison on Small Problems}
    \label{GCPP small comparison}
    \begin{tabular}{*7c}
        \toprule
        & \multicolumn{6}{c}{Solution Quality: \% $\leq$ ---\% above SA} \\
        \cmidrule(lr){2-7}
        ~ & -10 & -5 & 0  & 10 & 25 \\    
        \midrule
        \textbf{classical} & ~  & ~ & ~ & ~ \\    
        \midrule
        greedy & 0 & 0 & 6.9 & 6.9 & 8.3\\
        tabu & 1.4 & 1.4 & 88 & 92 & 96\\
        greedy tabu & 0 & 0 & 89 & 93 & 96 \\
        \midrule
        \textbf{quantum} & ~  & ~ & ~ & ~ \\    
        \midrule
        Advantage & 0 & 0 & 19 & 19 & 22\\
        \midrule
        \textbf{hybrid} & ~  & ~ & ~ & ~ \\    
        \midrule
        greedy Advantage & 0 & 0 & 68 & 69 & 74\\
        2000Q qbsolv & 0 & 0 & 75 & 79 & 85\\
        Advantage qbsolv & 0 & 0 & 81 & 81 & 88\\
        \bottomrule
    \end{tabular}
\end{table}

 Let us start with the small problems. When comparing the small problems for the General CPP in Table \ref{GCPP small comparison}, we see once again that greedy and Advantage by themselves do not perform well, but that together get strong results. For these small problems in the $0-250$ variable range it appears that the classical algorithms like SA and tabu are more effective than the quantum/hybrid approaches running on current D-Wave LEAP resources.
 
   \begin{table}
    \centering
    \caption{Comparison on Medium Problems}
    \label{GCPP medium comparison}
    \begin{tabular}{*9c}
        \toprule
        & \multicolumn{8}{c}{Solution Quality: \% $\leq$ ---\% above SA} \\
        \cmidrule(lr){2-9}
        ~ & -35 & -20 & -10 & -5 & 0  & 10 & 25 \\    
        \midrule
        \textbf{classical} & ~  & ~ & ~ & ~ \\    
        \midrule
        greedy & 0 & 0 & 0 & 0 & 0 & 0 & 0\\
        tabu & 0 & 0 & 21 & 37 & 42 & 63 & 95\\
        greedy tabu & 0 & 0 & 26 & 26 & 47 & 63 & 89 \\
        \midrule
        \textbf{hybrid} & ~  & ~ & ~ & ~ \\    
        \midrule
        2000Q qbsolv & 5.3 & 5.3 & 21 & 26 & 47 & 53 & 84\\
        Advantage qbsolv & 0 & 11 & 21 & 26 & 42 & 68 & 89\\
        \bottomrule
    \end{tabular}
\end{table}

When looking at the data for medium sized problems in the range of $250-1000$ variables in Table \ref{GCPP medium comparison}, we see that the hybrid methods are comparable in solution quality to the classical methods and will sometimes get higher quality results.

   \begin{table}
    \centering
    \caption{Comparison on Large Problems}
    \label{GCPP large comparison}
    \begin{tabular}{*9c}
        \toprule
        & \multicolumn{8}{c}{Solution Quality: \% $\leq$ ---\% above greedy tabu} \\
        \cmidrule(lr){2-9}
        ~ & -99.5 & -75 & -50 & -20 & 0 & 10 & 25 \\    
        \midrule
        \textbf{classical} & ~  & ~ & ~ & ~ \\    
        \midrule
        greedy & 0 & 0 & 0 & 0 & 0 & 0 & 0\\
        tabu & 0 & 0 & 0 & 4 & 20 & 32 & 36\\
        \midrule
        \textbf{hybrid} & ~  & ~ & ~ & ~ \\    
        \midrule
        2000Q qbsolv & 4 & 8 & 28 & 36 & 52 & 56 & 56\\
        Advantage qbsolv & 4 & 12 & 40 & 40 & 44 & 56 & 56\\
        \bottomrule
    \end{tabular}
\end{table}

Once we start to look at larger problems, as shown in Table \ref{GCPP large comparison}, we see that the quantum-classical hybrids start to significantly outperform the classical methods in terms of solution quality.

\section{Discussion}
\label{discussion}

The following insights come from the data in Section \ref{results} and intuition gained from the implementation of the Closed Undirected CPP and  variants \ref{undirected variant} through \ref{windy variant} in the generalized algorithm for the CPP. The observations come from running the algorithms both directly on the D-Wave 2000Q  and Advantage chip, and via the quantum-classical implementation of \emph{qbsolv}~\cite{Booth} on the QAs.

The differences between the original algorithm~\cite{Siloi} and our modified version 
are as follows: First, we use $x_{i,j}$ to represent the same binary variable as $x_{j,i}$ whereas the original version
treated them as two separate variables. The advantage of this approach is that it halves the number of variables used. Second, 
we are able to remove a now unnecessary constraint from the equation. The advantages of this are two-fold. One, removing the constraint makes understanding the QUBO and implementing it easier. Two, the removal of the second constraint 
reduces the QUBO's variable connectivity, allowing for larger problems to fit directly on the hardware. These changes lead to being able to run a 12 odd degree vertices problem directly on the 2000Q versus the previous 8 odd degree vertices.

Now let us talk about the generalized CPP algorithm. While handling a much more general class of problems, this algorithm can use a large number of variables. Depending on which variants are used, the number of variables can grow quadratically with the number of edges in the graph. The bright side is that the variables used are not in general fully connected. This means more variables may be used when running the problem on quantum hardware. For example, on the 2000Q D-Wave chip, 109 variables for the algorithm were successfully embedded on the hardware compared to the 64 variable maximum 
when fully connected. 

There are many ways the choice of variants can increase or decrease the QA efficiency. 
Specifying the start and/or end vertex will decrease the number of variables and somewhat decrease the connectivity between the variables.
When implementing a Rural Postman Problem, requiring fewer edges can greatly decrease the connectivity between the variables.
Variants \ref{k variant} through \ref{hierachy variant} all greatly increase the number of variables required and/or increase the connectivity between variables.

One of the main determining factors one has control of which affects how many variables are required is $i_{\text{max}}$, the maximum length of the walk allowed. To find a minimal walk weight which meets all 
criteria, one must allow sufficient steps in the walk to find that minimal walk weight. Roughly $(2\vert U\vert + \vert D\vert)i_{\text{max}}$ variables are required for all variants except \ref{k variant}, \ref{service variant}, and \ref{hierachy variant}, which require 
approximately some integer multiple more variables. Thus we 
try to pick a 
minimal, yet sufficiently large $i_{\text{max}}$. A safe value to pick, in the sense it will be 
sufficiently large for any variant, is $i_{\text{max}} = 2\vert E\vert$. If this is too many variables, one may 
try a smaller $i_{\text{max}}$. It is safer to greatly decrease $i_{\text{max}}$ from $2\vert E\vert$ when either there are a large number of undirected edges or when a significant number of edges are not 
required in the Rural Postman variant.  

Now let us take a moment to talk about the '$P$' variables from earlier, the ones which we multiply each constraint by when adding to our QUBOs. This is where our effort
becomes a bit more of an art than a science. From a mathematical perspective, one should choose the '$P$' variables to be arbitrarily large. From an implementation perspective this should not be done. When the QUBO is embedded on the hardware, all the values are scaled to fit within a specific range 
with limited precision and as such, if the '$P$' variables are chosen too large, then numbers which are 
not zero may be treated as zero, leading to poor results. There are some general guidelines for the 
choices. All the '$P$' variables are multiplied with constraints which, if broken, lead to an invalid solution. 
The '$P$' variables should at the very least be larger than the highest weight edge. The authors often found having all such
variables set between 1.5 to 15 times the highest edge weight worked well. If one tries to implement the
algorithms in this paper and gets results which lead to invalid solutions, then the likely culprit is the '$P$' variables. In this case, one should increase the '$P$' value for the constraint which is 
broken. If however one is getting valid, but non-optimal results, this may be caused by having '$P$' 
variables which are too large and one should try decreasing all of them slightly.

One of the surprising results 
is how effective combining annealing on a QA and greedy were, even when either 
method alone achieved poor results. An 
interpretation of why this occurs is as 
follows. The energy landscape for our QUBOs, especially the larger ones, is complex with 
many peaks and valleys of varying heights and depths. The greedy algorithm by itself can 
only ever go down, and so will descend into 
the nearest valley which has a low likelihood
of being the deepest valley or even a deep 
valley. When annealing on a QA, their is a 
strong likelihood of arriving at the deepest,
or at least one of the deepest valleys, but 
due to noise and flux errors \cite{DwaveQPUSolver}. 
has trouble settling to the bottom 
of these values. So when we combine these 
methods together, the QA finds one of the 
deepest valleys and then greedy quickly gets 
us to the bottom of the valley. 

Another surprising result appears in section \ref{GCPP Comparison Study}. For the methods tested, the data shows a comparative advantage for classical algorithms on small problems, but as the problems grow in size, the quantum-classical hybrid methods overtake the classical algorithms and achieve superior results. This trend is highlighted in Tables \ref{GCPP small comparison}, \ref{GCPP medium comparison}, and \ref{GCPP large comparison}.

One should note that in this paper we have defined our graphs to not include multi-graphs, graphs which may have more than one edge which go from vertex $i$ to vertex $j$. This is to make the notation simpler. Everything in this paper may be extended to work with multi-graphs with the largest obstacle being the notation. For ideas on how to implement this work for multi-graphs one should look at the QUBOs for variant \ref{k variant} and variant \ref{service variant}.

In conclusion, the authors have designed and developed a framework for solving a large number of variants of the CPP on a QA. 
Implementation of the framework for variants \ref{undirected variant} through \ref{windy variant} on the  D-Wave 2000Q were successful. Optimal results
were achieved for problems which could be embedded on the hardware with only short chains and optimal 
results were sometimes achieved for larger problems after 
tuning the '$P$' variables. 
Future directions include the following.
Implementation of the remaining variants 
outlined. Implementation of further variants as there are more variants which could be easily adapted to the method defined in sections \ref{BGC} and \ref{even more general}, but were not included to keep this paper reasonable in length. 
Translating the CPP algorithm and variants for gate-based quantum
architectures.
Developing a more efficient way to choose optimal '$P$' 
variable values given the inputs from the problem. Additionally, there is room to experiment with this 
algorithm in conjunction with an iterative and/or graph partitioning approach to the CPP.

\section*{Acknowledgements}
We acknowledge the ASC program at LANL for use of their Ising D-Wave 2000Q quantum computing resource. We also acknowledge the use of the D-Wave Leap 2000Q and Advantage quantum computing resources. Assigned: Los Alamos Unclassified Report LA-UR-22-27468.

\section*{Funding} This research was supported by the U.S. Department of Energy (DOE) National Nuclear Security Administration (NNSA) Advanced Simulation and Computing (ASC) program at Los Alamos National Laboratory (LANL). This research has been funded by the LANL Laboratory Directed Research and Development (LDRD) under project number 20200056DR. JEP, CFAN,
and SMM were funded by LANL LDRD. JEP was also funded by the U.S. Department of Energy (DOE) through a quantum computing program sponsored by the Los Alamos National Laboratory (LANL) Information Science \& Technology Institute. Assigned: Los Alamos Unclassified Report LA-UR-22-27468. LANL is operated by Triad National Security, LLC, for the National Nuclear Security Administration of U.S. Department of Energy (Contract No. 89233218NCA000001). The funders had no role in study design, data collection and analysis, decision to publish, or preparation of the manuscript.

\section*{Author Information}
\subsection*{Author Names and Affiliations}
\textbf{Computer, Computational, and Statistical Sciences Division, Los Alamos National Laboratory, NM, USA} \\
\textbf{Mathematics Department, University of California, Santa
Barbara, CA, USA} \\
Joel E. Pion \\
\textbf{Theoretical Division, Los Alamos National Laboratory, Los Alamos, NM, USA} \\
Christian F. A. Negre \\
\textbf{Computer, Computational, and Statistical Sciences Division, Los Alamos National Laboratory, Los Alamos, NM, USA} \\
Susan M. Mniszewski

\subsection*{Author Contributions}
J.E.P. and S.M.M. designed the project. J.E.P. performed the numerical simulations and optimizations. S.M.M. supervised the whole project. C.F.A.N advised on the mathematical formulations. All authors contributed to the discussion, analysis of the results and the writing of the manuscript.

\subsection*{Corresponding author}
Correspondence to Susan M. Mniszewski

\section*{Ethics declarations}
\subsection*{Declarations}
This work does not involve human participants and presents no ethical concerns.

\subsection*{Conflict of interest}
The authors declare no competing interests.

\subsection*{Human and Animal Ethics}
Not Applicable

\section*{Consent for publication}
All authors agreed to publication of this research.

\section*{Availability of data and materials}
All author-produced code will be available upon reasonable request.


\nocite{*}



\end{document}